\documentclass[twocolumn,pra,showpacs,multicol,amssymb]{revtex4}
\usepackage{graphicx}% Include figure files
\usepackage{dcolumn}% Align table columns on decimal point
\usepackage{bm}% bold math
\usepackage{graphics}
\usepackage{epsfig,color}
\usepackage{tabu}
\usepackage{array}

\newcommand{\be}{\begin{equation}}
\newcommand{\ee}{\end{equation}}
\newcommand{\bea}{\begin{eqnarray}}
\newcommand{\eea}{\end{eqnarray}}

\DeclareUnicodeCharacter{2212}{-}

\begin{document}
%\pacs{03.65.−w; 03.65.Vf; 03.67.Bg;75.10.Pq}
\title{Amplifying quantum correlations with quench dynamics in a quantum spin chain: Steady-states versus ground-states} 
\author{Sasan Kheiri}
\affiliation{Department of Physics, University of Guilan, 41335-1914, Rasht, Iran}
%\orcid{0000-0002-2445-2701}
\author{Hadi Cheraghi}
\email{h.cheraghi1986@gmail.com}
%\orcid{0000-0001-5885-0102}
\affiliation{Institute of Physics, M. Curie-Sk\l odowska University, 20-031 Lublin, Poland}
\author{Saeed Mahdavifar}
\email{smahdavifar@gmail.com}
\affiliation{Department of Physics, University of Guilan, 41335-1914, Rasht, Iran}
%\orcid{0000-0003-1985-4623}
\date{\today}

\begin{abstract}

We analyze the behavior of steady-state quantum correlations (QCs) in the spin-1/2 transverse field XY chains analytically, in terms of quench dynamics at zero-temperature. We show that  steady-state QCs are strikingly greater than the equilibrium ones in its ground-state, where a single quench is performed from ferro- into paramagnetic phases. Another framework to amplify the QCs here is a feasible protocol called double quench dynamics. To fulfill this purpose, we probe a middle quench point and spending time T (defined as the time passing from the middle quench point before reaching a second quench). By doing so, both single and double quenches act as practical tools to control the enhancement of the steady-state QCs in the final quenched point. In particular, and in parallel to
expectations for some other quantities, we indicate that the nonequilibrium quantum phase transitions also can be identified by nonanalyticities in the steady-state QCs. Our work may be testable with the current class of trapped-ion or ultracold-atom experiments, and encourage the possible potential in the quantum information field.

\end{abstract}
\maketitle

%%%%%%%%%%%%%%%%%%%%%%%%%%     Section I      %%%%%%%%%%%%%%%%%%%%%%%%%%%%%%%%%%%%%%%%%%%%%%%%%%%%%%%%%%%%%%%%%%%%%%%%%%%%%%%%%%%%%%%%%%%%%%%%%%%%%%%%%%%%%%
%%%%%%%%%%%%%%%%%%%%%%%%%%%%%%%%%%%%%%%%%%%%%%%%%%%%%%%%%%%%%%%%%%%%%%%%%%

\section{Introduction}\label{sec1} 

Quantum correlations (QCs) have become central for the characterization and the classification of  quantum matters both in- and out-of-equilibrium \cite{1, 2, 3, 4, 5}.
Exotic quantum phases such as spin liquids \cite{6,7}, topological \cite{8,9,9b}, and many-body localized systems \cite{10} have been manifested through QCs. 
It is demonstrated that entanglement in quantum many-body systems can be accessible in experiments such as in full-state tomography \cite{13,14} and  ultra-cold atoms to measure Renyi entropies \cite{15,16} although these experimental efforts are limited to few-particle systems.  Recently, using  inelastic neutron scattering,  an experimental protocol has been introduced to detect and quantify entanglement in the solid-state by  performing measurements on $\rm{Cs_2CoCl_4}$, described as a quasi one-dimensional spin-1/2 XXZ model with a transverse  field (TF) \cite{22b}.  

A growing interest in QCs $-$ especially bipartite sites $-$ has been intensified by the potential applications in information processing; for instance,  continuous quantum machines \cite{22c}, masking quantum information \cite{22d}, quantum networks \cite{22e}, and teleportation \cite{22f}. 
 Due to the novelty of this topic, there are highly complicated problems to tackle to let us unravel the mystery of such systems. A crucial open proposal, here, is that to know how to set up a state  (approximately) in equilibrium with a given Hamiltonian to gain  more QCs than those in its ground state. In this paper, we strive hard to tackle this problem in the framework of the nonequilibrium dynamics after a long-time evolution.
 
Non-equilibrium dynamic  system compared to systems in equilibrium, is a far more complex topic.   Therefore,  quantum  quenches provide a practical platform to examine the behavior of a system out of equilibrium \cite{23,24,25,26}. 
The quenches can be global, local, or geometrical, i.e., an intermediate between a local and a global quench \cite{26b}.
The essence of a quench outstandingly comes back to the different excess energy density measured with respect to the ground state energy of the post quench Hamiltonian. To clarify, in the global quenches, it remains finite while in the local quenches, when the system size increases, it vanishes. Recently, dynamical phase transitions in many body systems initially prepared in the state far-from-equilibrium have aroused considerable interest.
 
In the notation of the quantum quench, the concept of a \textit{steady-state} transition describes a nonequilibrium quantum phase transition. In a sense, this transition is triggered by a nonanalytic change in physical properties as a function of the quench parameter in the asymptotic long-time average of the system  \cite{27b}.  
Remarkably, the steady-state QCs  shows a significant increase (a super-Heisenberg scaling) in comparison with those in the ground state \cite{27c}.
This can be also applied to the dynamical quantum phase transition to characterize nonequilibrium criticality through the relationship of the singularities of Loschmidt echo to the zeros of local order parameters \cite{27d},  the rate function and the  fidelity susceptibility \cite{27e}, the energy gap \cite{27f}, and the delineation of the dynamical phase diagram of spin chains with long-range interactions \cite{27i}. 

In this paper, we are interested to know in what ways quench dynamics are more likely to affect steady-state QCs \cite{27j}.  
By employing concurrence  and quantum discord (QD) as measures QCs in the system, we extract the features of the steady-state QCs after sudden quantum quenches at zero-temperature. The most striking fact that has been shown here is that this will lead
to control and increase of  QCs in the thermodynamic limit in the quantum spin-1/2 chains.
This is more likely to allow us to identify quantum materials suitable for new applications  and give us novel insights into complex quantum phenomena.
Here, we focus on the integrable spin-1/2 XY chain model in a TF.  It should be noted that our results can be applied to  
the other integrable and non-integrable models. The equilibrium phase diagram of the model at zero temperature is comprised of two phases, gapped ferromagnetic (FM) 
and paramagnetic (PM) phases  separated at the critical field where a quantum phase transition occurs.  
We indicate that the quenching system from FM into PM regions creates a significant amount of QCs between nearest-neighbor pair spins in the long-time average. Moreover, a system exposed to the double quench provides the steady-state QCs at the final quenched point compared to  both in equilibrium at $t=0$ and in  the steady-state of a single quench. 
In our setups, we make an effort to show the amplification of  the steady-state QCs of system at the final quenched point $\lambda_f$ ($\lambda$ is control parameter) in  phase $C$ when it is initially placed at its ground state at  $\lambda_i$ in  phase $A$. Here, to provide a double quench before the system to be quenched  to $\lambda_f$ , we seek an appropriate middle quenched point  $\lambda_m$ in phase B and a spending time T on which the system remains to control the values of the steady- state QCs at the final quenched point. Phases $A$, $B$, and $C$ can be the same or not.  We reasonably infer that by a proper protocol, a double quench can create more  steady-state  QCs than those in the single quench. More importantly, we find that the steady-state QCs $-$ for both single and double quenches $-$ disclose a singularity of the nonequilibrium quantum phase transition exactly at the quantum critical point, provided that the QCs do not undergo a sudden death in the vicinity of the critical point.

%%%%%%%%%%%%%%%%%%%%%%%%%%%%%%%%%%%%%%%%%%%%
%%%%%%%%%%%%%%%%%%%%%%%%%%%%%%%%%%%%%%%%%%%%
%%%%%%%%%%%%%%%%%%%%%%%%%%%%%%%%%%%%%%%%%%%%
\section{The model}
We consider the Hamiltonian of the one-dimensional (1D) spin-1/2 XY  model in the presence of a TF as
\begin{eqnarray}\label{eq1}
{\cal H} &=& -J\sum\limits_{n = 1}^N {\left[ {(1 + \delta )S _n^x S _{n + 1}^x + (1 - \delta ) S _n^y S _{n + 1}^y} \right]} \nonumber\\
&-& h\sum\limits_{n = 1}^N {S _n^z}~,
\end{eqnarray}
where $S _n$ is the spin operator on the $n$-th site. $J>0$ denotes the FM exchange coupling.  $\delta$ and $h$ are the anisotropy parameter and the homogeneous TF, respectively.  $N$ is the system size (or number of spins) and we consider the  periodic boundary condition $S _{N+1}^\mu=S _1^\mu $ ($\mu=x,y,z$).
The model exhibits a quantum phase transition at  $h_c = J$ from  the FM phase ( $h < J$) to the PM phase ($h>J$) \cite{37}. 
It can be seen that on the circle $h^2+(J \delta) ^2=J^2$, the wave function of the ground state is factorized into a product of single spin states \cite{38} that makes different regimes in the phase diagram of the model in the view of the revivals for Loschmidt echo \cite{38b}.

The Hamiltonian is integrable and it can be mapped to a system of free fermions. This model is exactly solvable. By applying the Jordan-Wigner transformation \cite{39}, spins are mapped onto the one-dimensional noninteracting spinless fermions with the creation and annihilation operator,
\begin{eqnarray}\label{eq2}
{\cal H} &=& -\frac{J}{2}  \sum\limits_{n = 1}^N \left[ a_n^\dag a_{n + 1}+\delta a_n^\dag a_{n + 1}^\dag  + h.c. \right] \nonumber\\
& -& h\sum\limits_{n = 1}^N {\left( {a_n^\dag {a_n} - 1/2} \right)},
\end{eqnarray}
where $a_n$ is the fermionic operator. Performing a Fourier transformation as ${a_n} = (1/\sqrt N)\sum\nolimits_k {{e^{ - ikn}}}~ {a_k}$ makes the Hamiltonian as ${\cal H} = \sum\nolimits_{k > 0} {H_k} $ where $ H_k = {\cal A}_k(a_k^\dag {a_k} + a_{ - k}^\dag {a_{ - k}}) + i{\cal B}_k(a_k^\dag a_{ - k}^\dag  + {a_k}{a_{ - k}})$. A four-dimensional Hilbert space is spanned by the basis vectors $\left| 0 \right\rangle ,\left| k \right\rangle  = a_k^\dag \left| 0 \right\rangle ,\left| { - k} \right\rangle  = a_{ - k}^\dag \left| 0 \right\rangle$, and $\left| {k, - k} \right\rangle  = a_k^\dag a_{ - k}^\dag \left| 0 \right\rangle $ where $\left| 0 \right\rangle$  is the vacuum state. Only bilinear terms like $a_k^\dag a_{ - k}^\dag$ ensures that the parity of the total number of fermions provided by $n_k=a_k^\dag a_k+ a_{-k}^\dag a_{-k} $ is conserved for each value of $k>0$. Thus,
the states  $\left| 0 \right\rangle$ and $\left| {k, - k} \right\rangle $ are coupled to each other by the Hamiltonian, while the states $\left| k \right\rangle$  and $\left| -k \right\rangle $ remain invariant. Hence, in order to study the dynamics of the system, it is sufficient to project the Hamiltonian to the two dimensional subspace spanned by $\left| 0 \right\rangle$ and $\left| {k, - k} \right\rangle $ since the
ground state for each value of $k$ lies within this subspace. Using a Bogoliobov transformation presented by ${a_k} = \cos ({\theta _k})~{\alpha _k} + i\sin ({\theta _k})~\alpha _{ - k}^{\dag}$ leads to the quasiparticle diagonalized Hamiltonian as
\begin{equation}\label{eq3}
{\cal H} = \sum\limits_k {{\varepsilon }_k ( \alpha _k^\dag \alpha _k -1/2 )},
\end{equation}
where the energy spectrum is $\varepsilon_k = \sqrt{{\cal A}_k ^2+{\cal B}_k ^2}$ with
\begin{eqnarray}\label{eq4}
{\cal A}_k=-[J\cos(k)+h] ~~~~;~~~~{\cal B}_k= J \delta \sin(k),
\end{eqnarray}
and $\tan (2{\theta _k}) =- {\cal B}_k /{\cal A}_k $. For simplicity, we put $J=1.0$. Thus, the quantum critical line is $h_c=1.0$.

%%%%%%%%%%%%%%%%%%%%%%%%%%%%%%%%%%%%
%%%%%%%%%%%%%%%%%%%%%%%%%%%%%%%%%%%%
%%%%%%%%%%%%%%%%%%%%%%%%%%%%%%%%%%%%
\section{Setups and Tools}
Quantum many-body systems have been established to be a possible candidate for the implementation of quantum information protocols and powerful quantum computers \cite{1, 2, 3}. with tools such as the entanglement and the QD that are realized experimentally in various setups \cite{40,41,42}. 

Also, numerous studies have been extended within the scope of the non-equilibrium dynamics of QCs in many-body system. \cite{28,28b,29,30,31,32, 33,34,34b,35,36}. Take the entanglement entropy, the entanglement spectrum \cite{28,28b}, and the quantum mutual information \cite{29} as examples.
It should be noted that there is a connection between the Schmidt gap and order parameters \cite{30}. Moreover, the ratios of gaps of entanglement spectrum for a block of consecutive sites in the finite transverse field Ising chains are geared to obtaining universal information \cite{31}. 
In systems weakly coupled  to a dephasive Markovian environment, entanglement and quantum mutual information play a crucial role in destroying many-body localization \cite{32} and propagating QCs \cite{33}. Entanglement spectrum causes emergence of the dynamical quantum phase transition \cite{34}.
The entanglement entropy is exerted to examine the propagation of non-local QCs in a quench from the hermitian to the non-hermitian models \cite{34b}.
The most telling point in the study of the dynamics of the pairwise entanglement and the quantum discord (QD) in the spin-1/2 XXZ chain model \cite{35} is that contrary to the concurrence, QD emerges as a good identifier of a quench by revealing some nonanalytic behaviors when it exceeds critical points.
Further, multipartite entanglement turns out to be useful for the mentioned detection \cite{36}.

Having been confirmed the importance of the existence of QCs in executing quantum tasks, it is vital to seek quantum systems with enough QCs  especially at zero-temperature and approximately in equilibrium with a state under given circumstances in which the system reaches a steady-state.
To this end, with the use of the quench dynamics, we suggest a setup that provides more QCs in  spin-1/2 quantum chains. The most noticeable fact is that the larger quench gap necessarily does not produce the larger steady-state QCs. It crucially depends on how the setup launches. Accordingly, our results presented here are different from what has been reported for the entanglement entropy \cite{36b}.
In this work, by employing a single and a double quench together with the conception of steady-state, we change the nature of the system so that the particles of the many-body quantum system are forced to feel more QCs at the long-time average of the system.
Due to the rich structure of many-body states, a number of different QCs measures have been introduced.
Here, we examine the QCs in a bipartite system with QD and concurrence measures at zero-temperature. These measures quantify all correlations between two sites that reveal information about the  system \cite{1,2,3,22c,22d,22e,22f}.
%%%%%%%%%%%%%%%%%%%%%%%%%%%  Figure 1
\begin{figure}[t]
\centering{\includegraphics[width=60mm]{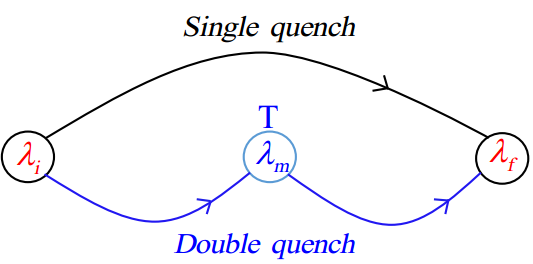}}
\caption{The schematic diagram of the setups.}
\label{fig:Fig1}
\end{figure}
%%%%%%%%%%%%%%%%%%%%%%%%%%%  Figure 1 

%%%%%%%%%%%%%%%%%%%%%%%%%%%%%%%%%%%%
%%%%%%%%%%%%%%%%%%%%%%%%%%%%%%%%%%%%
%%%%%%%%%%%%%%%%%%%%%%%%%%%%%%%%%%%%
\subsection{Setups}

We first fix the initial state of the system into the ground state of the initial Hamiltonian ${\cal H}(\lambda_i)$, where $\lambda$ indicates a control parameter that can be one or more.
  Then,  we use the idea of single quench and  double quench for the system as the following:\\
\\
(i) For single quench, at $t=0$, a quench starts from an initial state  (in   phase $A$) to a final state (in  phase $C$), as $\lambda_i$  to $\lambda_f$.  \\
\\
\\
(ii) For double quench, at $t=0$, a quench starts from an initial state  (in   phase $A$) to a middle state (in  phase $B$), as $\lambda_i$  to $\lambda_m$, and then a quench at $t=T$ from an evolved state (in   phase $B$) to a final state  (in  phase $C$), as $\lambda_m$ to $\lambda_f$. Phases $A$, $B$, and $C$ may be the same or not. 
\\
 The time-dependent Hamiltonian ${\cal H}(t)$ that models a double quantum quench is
\begin{eqnarray}\label{eq5}
{\cal H}(t) = \left\{ {\begin{array}{*{20}{c}}
{\begin{array}{*{20}{c}}
{{\cal H}(\lambda_i)}~,\\
{{\cal H}(\lambda_m)}~,\\
{{\cal H}(\lambda_f)}~,
\end{array}}&{\begin{array}{*{20}{c}}
{t \le 0}\\
{0 \le t \le T}\\
{t \ge T}
\end{array}}
\end{array}} \right.
\end{eqnarray}
with $\left| {\Psi (\lambda_i)}\right\rangle$ and $\left| {\Psi (\lambda_m)(T)} \right\rangle  = e^{ - iT{\cal H}(\lambda_m)}\left| {{\Psi (\lambda_i)}} \right\rangle $ that are the initial state at $t=0$ and the evolved state at $t=T$, respectively. The setups are schematically represented in Fig.~\ref{fig:Fig1}.
Accordingly, the system needs to be investigated in a long-time average at the final quenched point $\lambda_f$ where it fluctuates slowly  around a stable situation  at a steady-state. A system is in a steady-state if  the expectation values of local observable quantities become almost time-independent, by taking the limit $t  \to \infty $ \cite{27b,27c,27d,27e,27f,27i}. The nature of such a steady-state has been discussed in terms of the eigenstate thermalization hypothesis (ETH) \cite{t1,t2,t3,t4} that is beyond  the main theme of this work.
The main purpose of this paper is to indicate that this protocol helps us control and boost the values of the QCs at the long-time average of the system instead of QCs in the zero-temperature value at the final quenched point $\lambda_f$.
The control of amplifying steady-state QCs in the system will be attainable  in phase B by adjusting  the middle quench point $\lambda_m$ and the spending time $T$, defined as the time passing from the middle quench point before the system is exposed to a second quench.

%%%%%%%%%%%%%%%%%%%%%%%%%%%%%%%%%%%%
%%%%%%%%%%%%%%%%%%%%%%%%%%%%%%%%%%%%
%%%%%%%%%%%%%%%%%%%%%%%%%%%%%%%%%%%%
\subsection{Tools}.
\textit{Concurrence}.$-$ The concurrence is a measure of entanglement between two spins at sites $i$ and $j$ (here, we consider nearest-neighbors, $j=i+1$). It can be obtained from the corresponding reduced density matrix ${\rho _{i,i+1}}$ \cite{50}. In the fermionic picture, the reduced density matrix for two-point correlation functions can be written as
\begin{equation}\label{eq6}
{\rho _{i,i+1}} = \left( {\begin{array}{*{20}{c}}
{{X_{i,i + 1}^ +}}&0&0&{{-f_{i,i + 1}^*}}\\
0&{{Y_{i,i + 1}^ +}}&{{Z_{i,i + 1}^*}}&0\\
0&{{Z_{i,i + 1}}}&{{Y_{i,i + 1}^ -}}&0\\
{{f_{i,i + 1}}}&0&0&{{X_{i,i + 1}^ -}}
\end{array}} \right),
\end{equation}
where
\begin{eqnarray}\label{eq7}
X_{i,i+1}^{+}&=& \langle n_{i}n_{i+1}\rangle, \nonumber\\
Y_{i,i + 1}^ {+}  &=& \langle {{n_i}\left( {1 - {n_{i + 1}}} \right)} \rangle, \nonumber\\
Y_{i,i + 1}^ {-}  &=& \langle {{n_{i + 1}}\left( {1 - {n_i}} \right)} \rangle,  \nonumber\\
Z_{i,i+1}&=& \langle a_{i}^{\dag} a^{}_{i+1}\rangle, \nonumber\\
X_{i,i+1}^{-}&=& \langle 1-n_{i}-n_{i+1}+n_{i}n_{i+1}\rangle, \nonumber\\
f_{i,i + 1} &=&\langle a_{i}^{\dagger}  a_{i + 1}^{\dagger}\rangle, 
\end{eqnarray}
that $n_i={a_i^+}a_i$ is a fermionic occupation number of the $i$-th mode.
Therefore, the concurrence of the density matrix is given by
\begin{equation}\label{eq8}
C({\rho _{i,i + 1}}) = \mbox{Max} [ 0, \Lambda_{1}, \Lambda_{2} ],
\end{equation}
with 
\begin{eqnarray}\label{eq9}
\Lambda _ 1&=& 2 ( | Z_{i,i + 1} | -( X_{i,i + 1}^{+} X_{i,i + 1}^{-})^{1/2} ),  \nonumber\\
\Lambda_ 2&=& 2 ( | f_{i,i + 1} | -  ( Y_{i,i + 1}^{+} Y_{i,i + 1}^{-})^{1/2} ). 
\end{eqnarray}

\textit{Quantum discord}.$-$ To capture all QCs in a bipartite state that are not revealed by the concurrence \cite{51,51b}, one calculates the QD  \cite{52}. The QD is defined by the difference between total correlation, ${\cal I}(\rho _{i,i+1})$, and classical correlation, ${\cal C}(\rho _{i,i+1})$, as
\begin{eqnarray}\label{eq10}
QD_{i,i+1}=\mathcal{I}(\rho_{i,i+1})-\mathcal{C(}\rho_{i,i+1}).
\end{eqnarray}
To calculate the total and classical correlations, we use the reduced density matrix (eq.~({\color{blue}\ref{eq7}})) \cite{53}. In Appendix A, the derivation of the QD is explained in detail.

In the nonequilibrium dynamics of the system, the time-dependent two-point correlation functions are essential to derive the two sites reduced density matrix, given as 
\begin{eqnarray} 
R (t) =\frac{1}{N}\sum\limits_{l = 1}^N  \langle  a_{l}^{\dagger} a_{l+1} \rangle ~~;~~S(t)= \frac{1}{N}\sum\limits_{l= 1}^N  \langle a_{l}^{\dagger} a_{l+1}^{\dagger}\rangle    \nonumber
\end{eqnarray}
which for the first quench,it is given by
\begin{eqnarray}\label{eq11}
R (t) &=&\frac{1}{N}\sum\limits_{k>0} \cos(k) \Big[ 1- \cos(2\theta_k^{f}) \cos(2\Phi_k^{f})  \nonumber\\
&-&\sin(2\theta_k^{f})\sin (2{\Phi_k^{f}})\cos(2\varepsilon_k^{f}t)\Big],  \nonumber\\
S(t)&=&\frac{1 }{N}\sum\limits_{k > 0}\sin(k)\sin(2\Phi_k^{f})\Big[\frac{\sin(2\theta_k^{f})}{\tan(2\Phi_k^{f})}  \nonumber\\
&-& \cos(2\theta_k^{f}) \cos(2\varepsilon_k^{f}t)-i\sin(2\varepsilon_k^{f}t)\Big],  
\end{eqnarray}
with $\Phi_k^{f}= \theta _k^{f}- \theta _k^{i}$. 
\\ Also, for  the second quench  for $t \ge T$, it is as following
\begin{eqnarray}\label{eq12}
R (t)&=&   \frac{2}{N}\sum\limits_{k>0} \cos(k)|Q_k(t)|^2,   \nonumber\\
S(t)&=& - \frac{2}{N}\sum\limits_{k>0}  \sin(k) Q_k^\star (t)P_{ k}^\star(t), 
\end{eqnarray}
where
\begin{eqnarray}
P_k(t) &=& p_k(t)\cos (\Phi _k^{m}){e^{ - iT\varepsilon _k^{m}}}     \nonumber\\
&+& q_k(t)\sin (\Phi _k^{m}){e^{iT\varepsilon _k^{m}}}, \nonumber\\
Q_k(t)&=& q_k(t)\cos (\Phi _k^{m}){e^{iT\varepsilon _k^{m}}}     \nonumber\\
&-& p_k(t)\sin (\Phi _k^{m}){e^{ - iT\varepsilon _k^{m}}},  \nonumber
\end{eqnarray}
with
\begin{eqnarray}
p_k(t)&=& \cos (\theta _k^{f})\cos (\Phi _k^{f}){e^{ - i(t-T)\varepsilon _k^{f}}}  \nonumber\\
&+& \sin (\theta _k^{f})\sin (\Phi _k^{f}){e^{i(t-T)\varepsilon _k^{f}}}, \nonumber \\
q_k(t)&=&\sin (\theta _k^{f})\cos (\Phi _k^{f}){e^{i(t-T)\varepsilon _k^{f}}} \nonumber \\
&-& \cos (\theta _k^{f})\sin (\Phi _k^{f}){e^{ - i(t-T)\varepsilon _k^{f}}}, \nonumber
\end{eqnarray}
that we have used   $Q_{-k}(t)=-Q_k(t)$ and  $P_{-k}(t)=P_k(t)$. For the second quench, the quantity $\Phi_k^{f/m}= \theta _k^{f/m}- \theta _k^{m/i}$  is the difference between the Bogoliubov angles diagonalizing the pre-quench and the post-quench  Hamiltonians.  $\theta _k^{i}$ belongs  to the energy spectrum $\varepsilon _k^{i}$ with similar definitions for $\theta _k^{m}$  and $\theta _k^{f}$. 
%%%%%%%%%%%%%%%%%%%%%%%%%%  Figure 2
\begin{figure}[t]
\centering{\includegraphics[width=70mm]{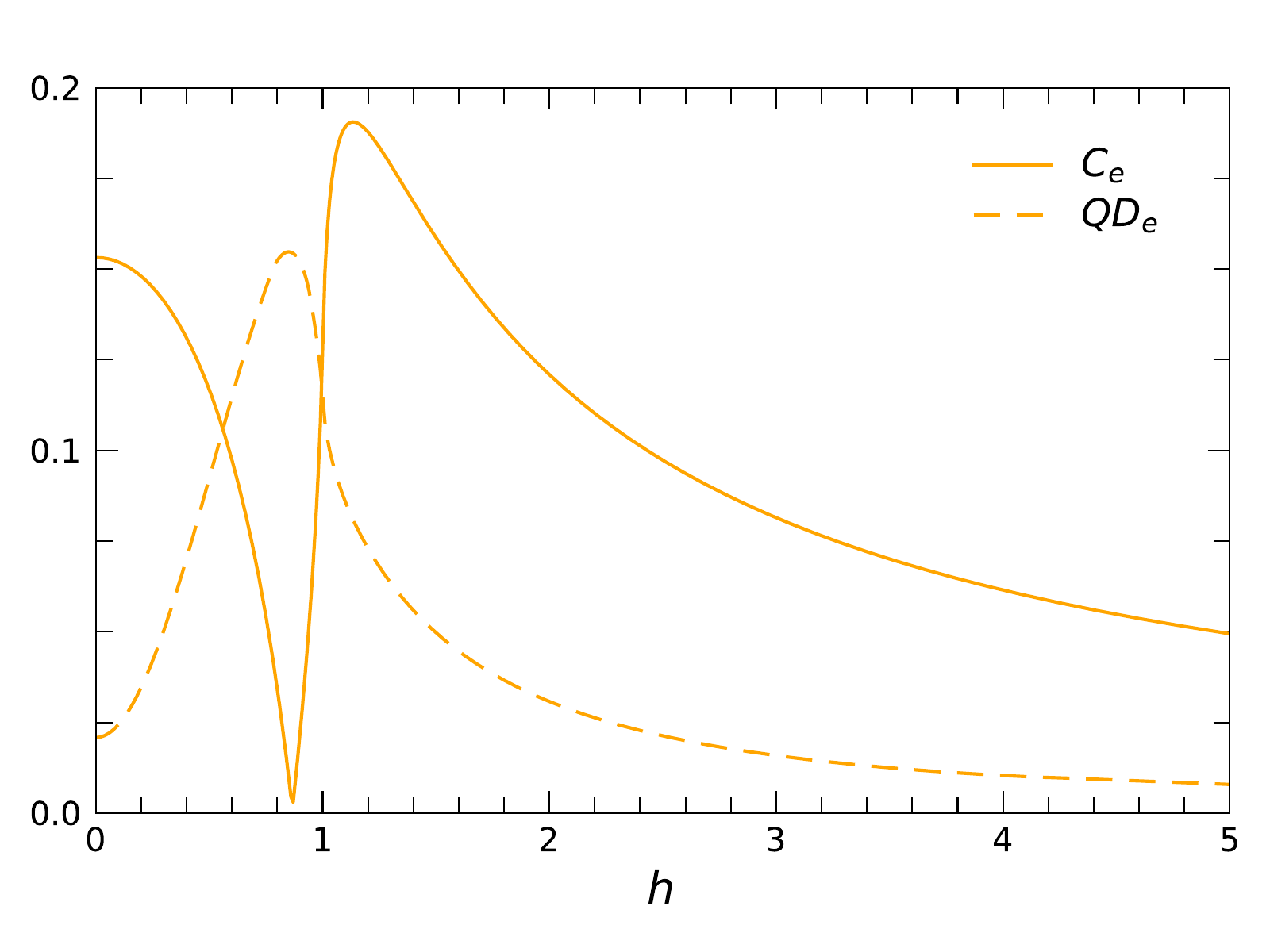}}
\caption{The equilibrium values of the concurrence (solid plot) and the QD (dashed plot)  at $t=0$ for the nearest-neighbor pair spins  as a function of the TF for $\delta=0.5$.}
\label{fig:Fig2}
\end{figure}
%%%%%%%%%%%%%%%%%%%%%%%%%%%  Figure 2

%%%%%%%%%%%%%%%%%%%%%%%%%%%%%%%%%%%%
%%%%%%%%%%%%%%%%%%%%%%%%%%%%%%%%%%%%
%%%%%%%%%%%%%%%%%%%%%%%%%%%%%%%%%%%%

\section{ Results and Discussions}
 We first should mention that before a quench in which the system is in an equilibrium situation, the QCs between nearest-neighbor pair spins in the ground state are known.  The graph in   Fig.~\ref{fig:Fig2} provides information about the equilibrium values  of the concurrence ($\rm{C_e}$) and the QD ($\rm{QD_e}$) for the nearest-neighbor pair spins  versus the TF for $\delta=0.5$. It is noticeable that at the factorized point, the concurrence is zero  while the QD reaches its maximum value \cite{53b}. With the increase of the TF,  the concurrence first grows swiftly, and peaks around the critical TF, and then falls exponentially. This pattern is the same for  QD. In the following, we argue about the steady-state QCs  for the nearest-neighbor pair spins in the quench dynamics of the system.
In this spirit, the long-time average of a parameter ${\cal O}$, ( as $\langle {\cal O} \rangle  = \mathop {\lim }\limits_{\tau  \to \infty } \frac{1}{\tau }\int_0^\tau  {{\cal O}(t)dt} $) lead to the system lies in the steady-state.
In the subsequent numerical plots without losing generality, we put a fixed anisotropy parameter as $\delta=0.5$.

%%%%%%%%%%%%%%%%%%%%%%%%%%%%%%%%%%%%
%%%%%%%%%%%%%%%%%%%%%%%%%%%%%%%%%%%%
\subsection{ Single quench}
\begin{figure}[t]
\centering{\includegraphics[width=83mm]{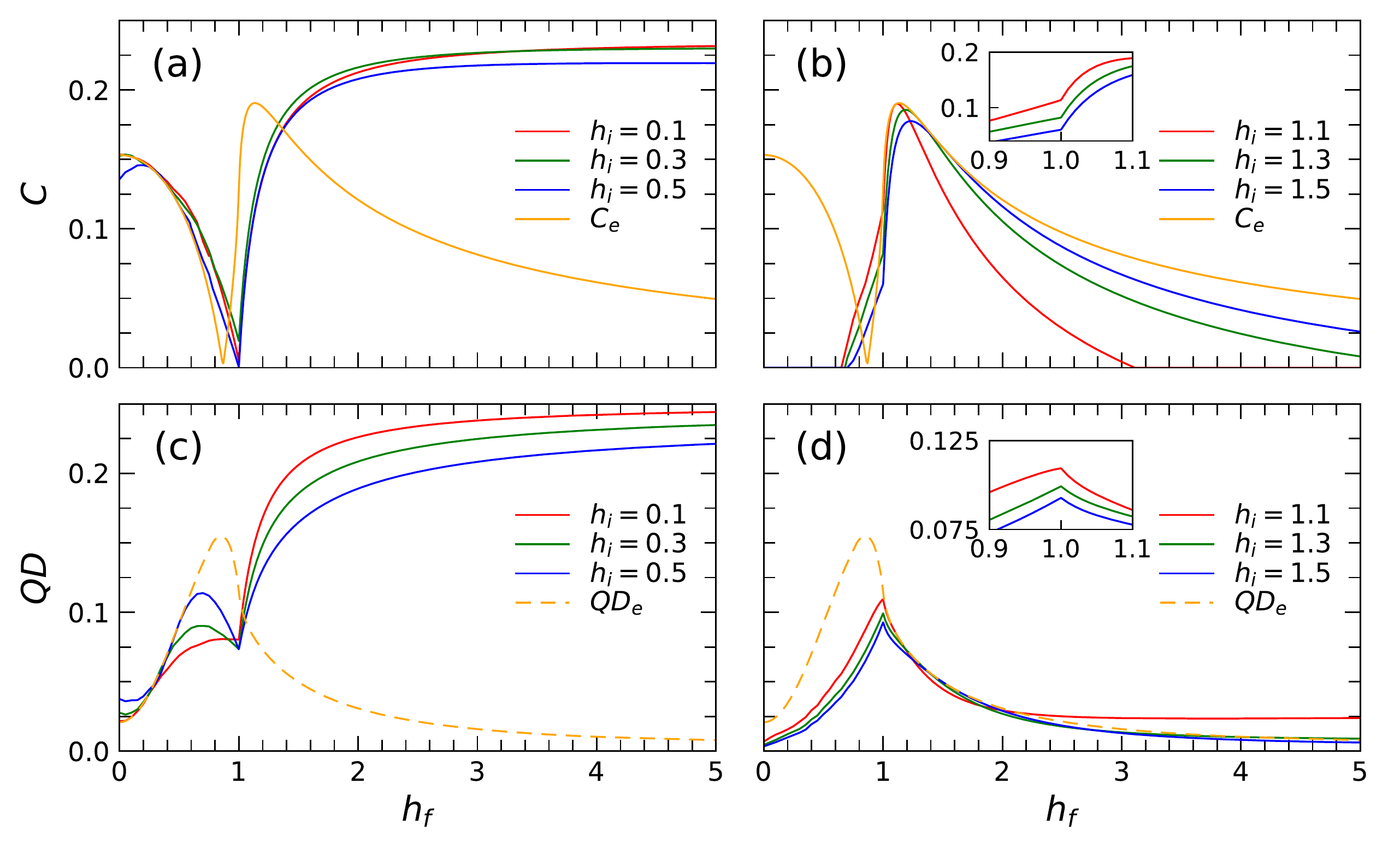}}
\caption{(color online) The steady-state QCs as a function of final TFs for quenches from (a), (c) the FM phase, and (b), (d) the PM phase. For more clarity of non-analytic behaviors of Figs.~(b) and (d) at the quantum critical point $h_c=1.0$, we have added their insets. }
\label{fig:Fig3}
\end{figure}

In a single quench, a system first is initially prepared in its ground state at $h_i$, where $h_i$ is the transverse field, then the TF is switched suddenly to a final value $h_f$. Now, using the time evolution operator, we seek the dynamic of the system.
 In Fig.~\ref{fig:Fig3}, we have shown the steady-state concurrence  and the steady-state QD versus the final TF.  Figures.~\ref{fig:Fig3}(a) and \ref{fig:Fig3}(c) correspond in the quenches  in the FM phase. As it can be seen, for quenches in the PM phase, at the long-time, both the concurrence  and the QD will share greater steady-state QCs between nearest-neighbor pair spins  compared to  QCs in equilibrium states. The reason behind this is  that,in the PM phase, the system in the equilibrium is almost polarized while the time-dependent physical state is a superposition of the ground and the excited states of the Hamiltonian at the final quenched point.  Figures.~\ref{fig:Fig3}(b) and \ref{fig:Fig3}(d) indicates the steady-state QCs versus final TF where quenches are in the PM phase. As a consequence, this process is unable to produce the QCs more than the system in the equilibrium. 
 
It should be mentioned that all quenches applied to the quantum critical point, $h_f=1.0$, lead to a steady-state QC with a nonanalyticity, displaying a nonequilibrium phase transition with $h_f$ as a control parameter. This is similar to the results for the model in which long-time average of the order parameters disclose nonequilibrium criticality when quenching crosses an equilibrium quantum critical point \cite{27d,27e,27f,27i,57}. However, in this present study, a nonanalyticity appears only when the final quenched point itself is in a critical point. Since the properties of the proposed nonequilibrium criticality in   Fig.~\ref{fig:Fig3}  are of interest, in Appendix B, we address it by making a comparison between the steady-state QCs and Loschmidt echo.

%%%%%%%%%%%%%%%%%%%%%%%%%%  Figure 3  
%%%%%%%%%%%%%%%%%%%%%%%%%%%  Figure 3 

 %%%%%%%%%%%%%%%%%%%%%%%%%%%  Figure 4
\begin{figure*}[t]
\centering{\includegraphics[width=160mm]{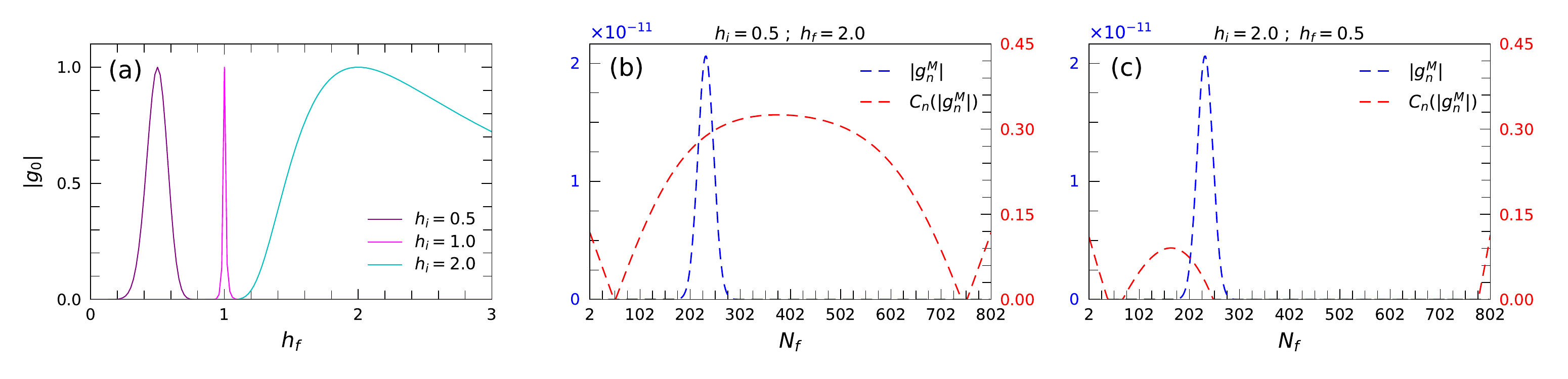}}
\caption{(color online) (a)  The expansion coefficient $|g_0|$ versus $h_f$ for quenches from $h_i=0.5,1.0,2.0$.  The maximum values of the expansion coefficients (for $n \ne 0$) and associated entanglements  are demonstrated  versus the occupation number $N_f$ for quenches from (b) $h_i=0.5$ to $h_f=2.0$, and (c) $h_i=2.0$ to $h_f=0.5$. Here, the system size is $N=800$. }
\label{fig:Fig4}
\end{figure*}
%%%%%%%%%%%%%%%%%%%%%%%%%%%  Figure 4

Now, we speculate the role of the excited states emerged from the dynamics of the system to create or annihilate the entanglement in long-time dynamics of the system. The time-dependent physical state of the system can be written as
$ | \Psi  (t) \rangle=\sum_{n=0} g_n e^{-i E_n^{f} t} | E_n^{f} \rangle$, where the eigenvalues $\{E_n^{f}\}$ and eigenvectors $\{   | E_n^{f} \rangle   \}$ 
with $n=0,1,2,...,$ corresponding to the Hamiltonian  at the quenched point. In addition,  the expansion coefficients are $g_n = \langle {E_n^{f} | \Psi (0)} \rangle $. The exact solution for the expansion coefficients is as 
%%%%%%%%%%%%%%%%%%%%%%%%%%%%%%%%
\begin{eqnarray} \label{eq13}
g_{n}=\prod\limits_{k > 0} \langle {E_n^{f}| [\cos (\Phi _k^{f})-i \sin(\Phi _k^{f}) \alpha _k^{\dagger f} \alpha _{-k}^{ \dagger f}] |    E_0^{f}} \rangle. \nonumber\\
\end{eqnarray}   
where $\alpha _k^f$ is related to the quasi-particle diagonalizing the quenched Hamiltonian.
The time-dependent concurrence can be written as
\begin{eqnarray}\label{eq14}
C (t) &=& \sum\nolimits_{n} |g_n|^2  C_{n} \nonumber\\
&+& \sum\nolimits_{n \ne n'} g_n^{\star} g_{n'} e^{-i (E_{n'}^{f}-E_n^{f}) t}  C_{n,n'}
\end{eqnarray}
with $C_{n,n'} =\langle E_n^{f}| C |E_{n'}^{f} \rangle $.
At long-time average, the steady-state  concurrence yields to
\begin{eqnarray}  \label{eq15}
\langle C \rangle = |{g_0}{|^2} C_{0}+\sum\nolimits_{n \ne 0} |g_n|^2 C_n,
\end{eqnarray}
where $C_n$ indicates the value of concurrence in the excited states of the quenched Hamiltonian. For the ground state, we have $n=0$ and  $g_0=\prod\nolimits_{k > 0}  \cos (\Phi _k^{f})$. 
As can be seen from eq.~(\ref{eq14}), the time-dependent concurrence is composed of two terms, one of which is constant and the other is time-dependent. By applying the long-time average condition, we extract eq.~(\ref{eq15}) that reveals the steady-state  concurrence.
It can be inferred that only this term plays a significant role in the steady-state concurrence.

A notable feature in eq.~(\ref{eq13}) is that only subspaces with an even-number of fermions should be explored. In other words, $g_n$ for an odd-number of fermions is zero.  In Fig.~\ref{fig:Fig4}(a), $|g_0|$ is plotted versus quenched TF for initial states in the FM, PM phase, and the critical point. It is clear that $|g_0|=1$, only happens when there is no quench. Another point to be noted is that since quenching from one phase into another breaks the symmetry of the ground state, $|g_0| $  tends to be zero. As a result, high excited states can contribute significantly to the steady-state concurrence.
 To seek more information about the role of the excited states, we have calculated the maximum values of the  expansion coefficient $|g_n^{M}|$  in  the different even-subspaces in the interval $k \in [0,\pi]$. By doing so, we detect a special wave number $k^*$. 
The results are illustrated for a quench from the FM to the PM phases   in (Fig.~\ref{fig:Fig4}(b))  and then from the PM to the FM phases in (Fig.~\ref{fig:Fig4}(c)) for a chain with size $N=800$ as a function of the occupation number $N_f$, as $N_f = \sum\nolimits_k \langle E_n^{f} | \alpha _k^{\dagger f} \alpha _k^{f}  | E_n^{f} \rangle$. It should be mentioned that $|g_n^{M}|$ is the same for both cases. 
Regions with a high number of fermions, it reached to the highest value in which the energy spectrum contains  high energy excited branches. Moreover, the concurrence belonging to $|g_n^{M}|$ reveals different behaviors. 
 As shown, its amount in the overlapping region, for panel (b) is more than panel (c).
The most noticeable point is that in the high energy excited branches, the pair spins for quenches from the FM to the PM  phases are  entangled in all subspaces  and they display a complete overlapping with $|g_n^{M}|$. That is why we witness a significant value of the entanglement ($\sum\nolimits_{n \ne 0} |g_n|^2 C_n$) that is considerably larger than its ground state value.
In contrast, in a quench from the PM to the FM phases, the overlap is restricted to a part (panel(c)). 
Consequently, that  similar to the former , where quench does not happen.

\begin{figure}[t]  
\centering{\includegraphics[width=83mm]{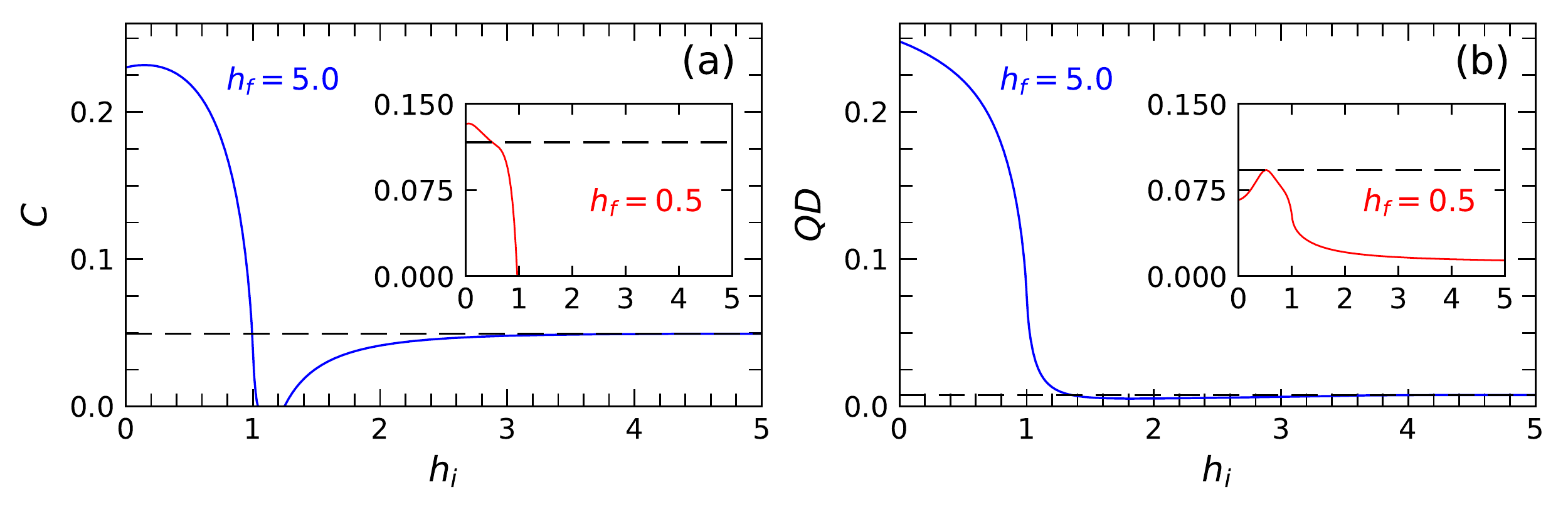}}
\caption{(color online)  The steady-state (a)  concurrence and (b)  QD as a function of initial TFs for quenches into the PM phase (the plots) and the FM phase (the insets), respectively. The horizontal black dashed lines indicates the QCs in  the equilibrium at $h_f$.}
\label{fig:Fig5}
\end{figure}

A deeper insight into the quench dynamics of the system is investigated by focusing the role of the initial state on the amount of the steady-state QCs at the final quenched point. 
The graphs, in Fig.~\ref{fig:Fig5}, compare figures for different values of h, specifically at $h_f=0.5, 5.0$.
 It is illustrated that when quenches is applies to the PM phase, 
with initial state in the FM phase, it leads to more QCs compared to the case in which the initial state is in the PM phase.
Moreover, the QCs for the initial states in the PM phase  far from the quantum critical point reveal a constant value close to the equilibrium, independent of the initial state though. For the QD, this value is almost zero . Another point to be mentioned is that in the PM phase, there is a region very close to $h_c=1$  where  the concurrence is zero. The pattern is reverse for the QD in which it is more than its equilibrium.
However, the insets of Fig.~\ref{fig:Fig5} are related to quenches from the desired initial states into a final quenched point in  the FM phase as $h_f=0.5$. Based on the exact results,  for an initial state within the PM phase, the pair spins will not entangle at long times. To put it more simply, the concurrence does not exist while the QD is minor. Also, only initial states within $h_i<h_f$ can have more entanglement values than their equilibriums. It should be noted that the initial states always generate the steady-state QD less than its equilibrium.

%%%%%%%%%%%%%%%%%%%%%%%%%%%  Figure 5 
%%%%%%%%%%%%%%%%%%%%%%%%%%%  Figure 5

%%%%%%%%%%%%%%%%%%%%%%%%%%%  Figure 6  
\begin{figure}[t] 
\centering{\includegraphics[width=83mm]{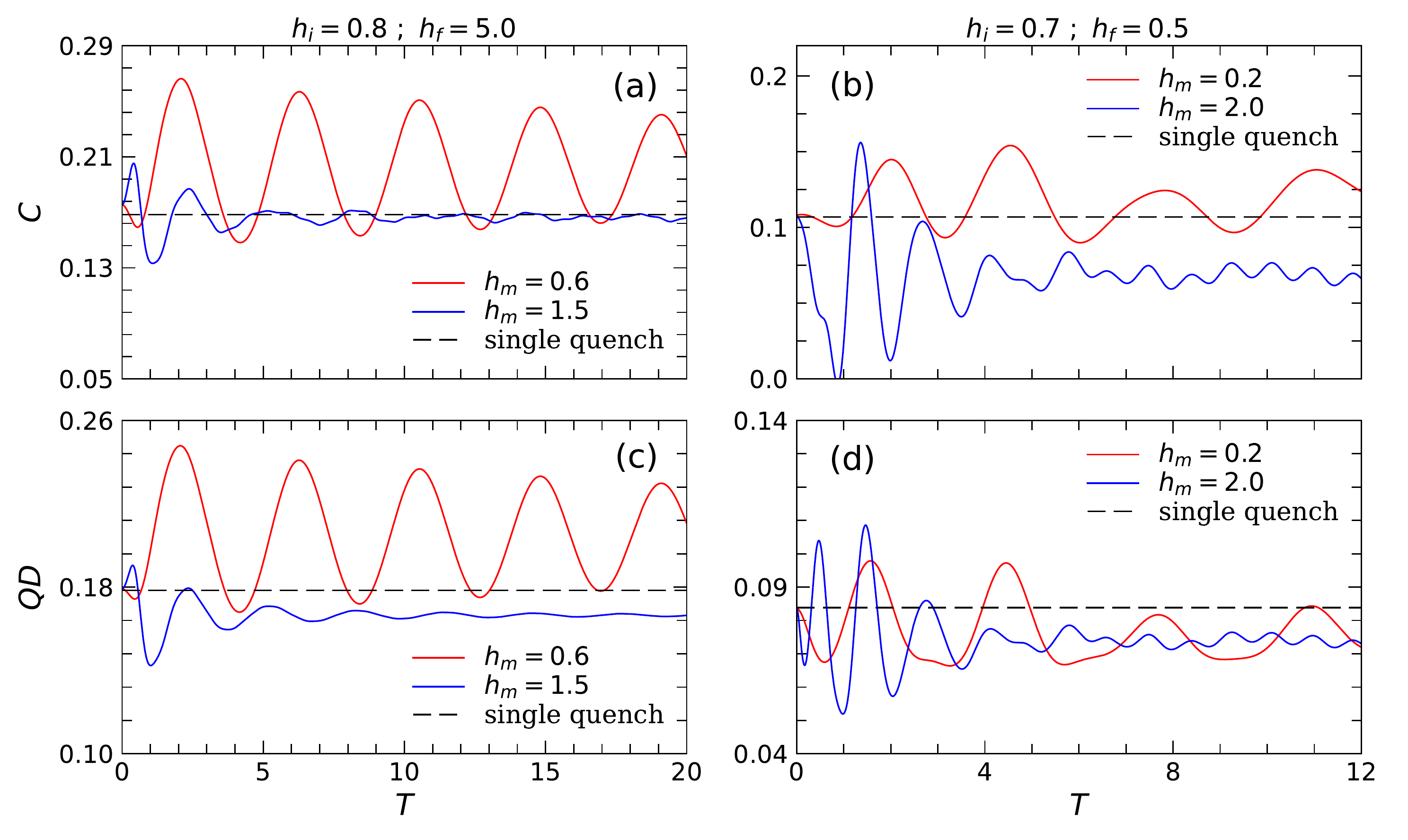}}
\caption{(color onlie) The  steady-state QCs versus the spending time $T$ for a quench from (a), (c) $h_i=0.8$ to $h_f=5.0$ and (b), (d) $h_i=0.7$ to $h_f=0.5$, where middle quenched points are situated at the FM and the  PM phases. The horizontal black dashed lines illustrates the  steady-state  QCs for the single quench case as $h_i \to h_f$.  }
 \label{fig:Fig6}
\end{figure}
%%%%%%%%%%%%%%%%%%%%%%%%%%%  Figure 6

%%%%%%%%%%%%%%%%%%%%%%%%%%%%%%%%%%%%
%%%%%%%%%%%%%%%%%%%%%%%%%%%%%%%%%%%%
\subsection{ Double quench}
To study the dynamics of the QCs in the double quench dynamics, we apply a middle quenched point $h_m$ before the system is exposed to a quench from the initial point $h_i$ to the final point $h_f$.
In the middle quenched point, the system will evolve up to the given time $t=T$. At this specific time, the system is exposed to the second quench and it continues to evolve into the final Hamiltonian. Both control parameters, $h_m$ and $T$ are responsible for controlling the value of the steady-state QCs in the final quenched point.  

In Fig.~\ref{fig:Fig6}, we plot the steady-state concurrence and the QD versus the spending time $T$ for the different middle points $h_m$. Also, the steady-state QCs for a single quench from $h_i$ to $h_f$ are shown by the horizontal black dashed lines. 

The main purpose of this work is to explore amplified values of the QCs at a final quenched point into both the FM and the PM phases as $h_f$ = 0.5; 5.0.  Here, the system initially is  placed in the FM phase. We elect two middle points; one in the FM phase, and another in the PM phase. As it is evident, by selecting out an appropriate middle point in the FM and the PM phase, some given spending times will appear at which the system can gain more QCs. The most intriguing fact is that, in some specific times, the QCs  for the second quench are exactly the same as  QCs in the single quench. This is common to protocols  in which the QCs in double quench cross QCs in  single quenches. It shows that neither $h_m$  nor spending time $T$ has an impact on steady-state QCs.

%%%%%%%%%%%%%%%%%%%%%%%%%%%  Figure 7 
\begin{figure}[t]
\centering{\includegraphics[width=80mm]{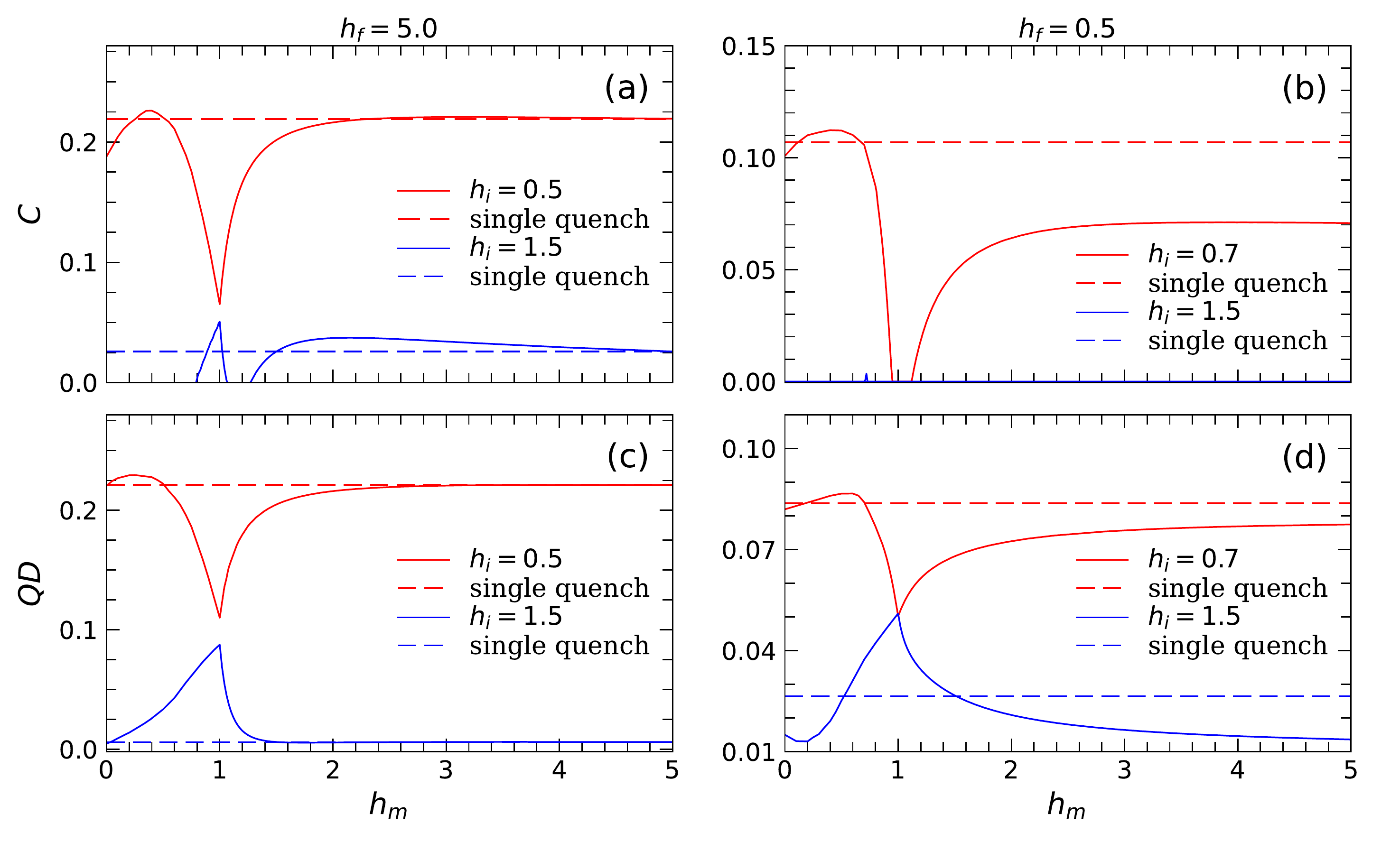}}
\caption{(color online) The  steady-state QCs as a function of  $h_m$  when the system is in a steady-state at the middle point for quenches from the FM and the  PM phases into final points $h_f=0.5, 5.0$. The horizontal black dashed lines illustrates the  steady-state  QCs for the single quench case as $h_i \to h_f$.  }
\label{fig:Fig7}
\end{figure}
%%%%%%%%%%%%%%%%%%%%%%%%%%%  Figure 7 

%%%%%%%%%%%%%%%%%%%%%%%%%%%  Figure 8 
\begin{figure}[t]
\centering{\includegraphics[width=83mm]{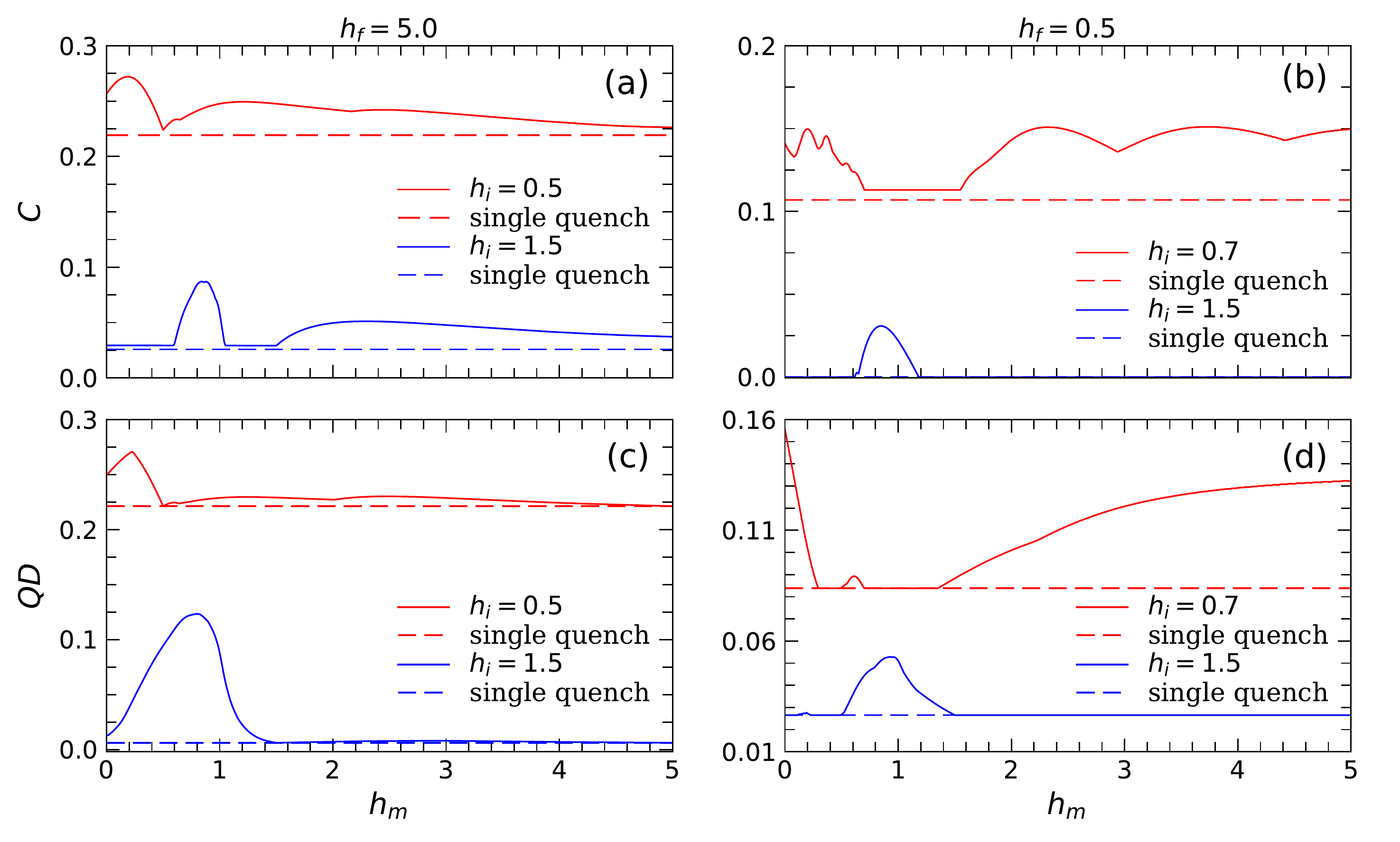}}
\caption{(color online) The same as Fig.~\ref{fig:Fig7} with the difference that here at the middle point, the time $T$ is chosen from the interval [0,10] in such a way that it belongs to  the maximum value of QCs. }
\label{fig:Fig8}
\end{figure}
%%%%%%%%%%%%%%%%%%%%%%%%%%%  Figure 8

A notable feature in Fig.~\ref{fig:Fig6} is that for the middle point, in the FM phase, the  oscillations spread on larger spending time  $T$ that is vividly illustrated in panels (red color). The steady-state QCs oscillate around a mean value  $-$ faster or slower depending on the type of quench $-$ and with a quench-dependent period $T_{\rm{osc}}$.
For the middle point in the PM phase, the pattern is the same while a small irregularity arises.  Moreover, it can be inferred from Fig.~\ref{fig:Fig6}(a), (c) and Fig.~\ref{fig:Fig6}(b), (d) that $T_{\rm{osc}}$ for a certain quench (e.g. $h_i$=0.8, $h_f$=0.5 ), the QCs in the PM phase coincide with those in the FM phase.
By Intuition, for the middle point in the PM phase, the steady-state QCs at the  middle point damp sharply at the short spending timing spend $T$ ((time measure shown in the Fig) (blue color). For these chosen quenches, panels (a)-(c) reveal that if the system stays in the middle point in the FM phase for a long-time, we can capture more steady-state QCs than those in the single quench. In contrast, this case never arises from the middle point in the PM phase.

Eventually, we  investigate how the middle point affects the steady-state QCs in the process of the double quench dynamics. 
First, we explore the behavior of QCs when the system reaches a steady-state situation in the middle point (Fig.~\ref{fig:Fig7}).
Fig.~\ref{fig:Fig7} shows changes in the steady-state QCs as a function of $h_m$  when the system is in a steady-state at the middle
point for quenches originated from the FM and the PM phases into final points $h_f$ = 0:5; 5:0.

Then, we examine the behavior of QCs,  when the system is exposed to the double quench at a specific time where the QCs have their maximum values. (Fig.~\ref{fig:Fig8}). In the latter case, we let the system evolves at the middle point up to $T=10$ and we elect the maximum of the QCs in this interval. So the  spending time  for the latter varies.
 The results demonstrated in Figs.~\ref{fig:Fig7} and \ref{fig:Fig8}  correspond to quenches from different initial and final states of the FM and the PM phases into desired middle points. 
 As shown in Fig.~\ref{fig:Fig7}, a steady-state situation in the middle point is unable to create remarkable values of the QCs, while we witness amplified QD at $h_i$  in the PM phase. 
The quantum critical point at $h_m$ in the steady-state situation unveil  an explicit signature by generating minimum and maximum values  and a  nonanalytic  behavior of a nonequilibrium phase transition exactly at the equilibrium critical point, provided that sudden death does not occur (please see Appendix B). 
 The quenching at a given time $T$ when the QCs achieve the maximum value at an optional interval $T=[0,10]$, makes the possibility of attaining more steady-state QCs in double quenches than those in single quenches (Fig.~\ref{fig:Fig8}).
As a consequence,  to have QCs  more than both the ground-state and the single quench  is possible by properly adjusting the middle point and timing spend $T$. To harness more QCs than those in the ground-state and the single quench  seems technically feasible in the double quench through  adjusting the middle point and spending time $T$.

%%%%%%%%%%%%%%%%%%%%%%%%%%%%%%%%%%%%
%%%%%%%%%%%%%%%%%%%%%%%%%%%%%%%%%%%%
%%%%%%%%%%%%%%%%%%%%%%%%%%%%%%%%%%%%
\section{ Conclusion}
Quantum information relies on operations beyond classical ones that nowadays experimentally is feasible by ultracold atoms and molecules in optical lattices  as the nearly ideal platform  for studying quantum many-body systems \cite{13,14,15}. To perform quantum tasks, the system under study must have a sufficient QC.
To gain the dramatic amount of QCs in a quantum system  that is not necessarily in its  ground state is one of the
challenges at the study of quantum processes.

We proposed a way in which a state is almost in equilibrium with the  system to provide more QCs.  This protocol facilitates the study of quantum procedures even in the systems with small QCs in their ground states.
This study is based on quantum quench, i,e., a simple protocol to bring a system out of equilibrium.
In fact, the goal is to generate the steady-state QCs by the quench dynamics in the long-time evolution.
Due to the exact solvability of several 1D quantum systems that represent an ideal framework corresponding to experimental data, in this paper, we  focused on the 1D spin-1/2 XY model in the presence of a TF. The motivation behind this paper is that the integrable 1D many-body models can be experimentally engineered and their nonequilibrium dynamics have been investigated in more detail in many contexts \cite{27,27j}.

We here established our setups upon single and double quantum quenches subjected by a TF,  and investigated the steady-state QCs in the final quenched points. It has been shown that how one can gain a more steady-state QCs than those in the ground state while we use concurrence and QD as measures of QCs.
A single quench from the FM into the PM phases showed up an outstanding amount of QCs while the same pattern is not seen from the PM to FM phases. To address  this issue, we put forward an analytical relationship for the steady-state concurrence and inferred that for the former case, there is an entirely overlap between the maximum values of the expansion coefficients and related concurrence while for the latter one, it is not seen. 

In the next step, we wonder if it is possible to increase the QCs more? To explore it, we used an interesting idea, a double quench, as a way that the system before locating in the final point will be quenched to a  middle point and we evaluated the system at this point for a given time $T$. Afterwards, at the time $T$, the system is exposed to the second quench. Spending time $T$ can be for a short time or for a long-time so that the system sets in a steady-state situation at the middle point. Furthermore, the middle point can be chosen in the FM or the PM phases. The aim is to select the middle point and spending time so that the system achieves a greater steady-state QCs at the final point compared to the single quench. Our results clearly demonstrated  that this can be feasible through the control of these parameters.

Additionally, it has been shown that nonequilibrium quantum phase transitions can be identified by non-analyticities in the long-time average
of some parameters such as order parameters, Loschmidt echo, and quantum Fisher information when the quenched parameter as the control parameter  crosses  the critical point. 
Here, we extended this topic to the QCs and proved that a nonanalyticity will emerge if the final magnetic field as a control parameter crosses  the critical point when the system is placed at a steady-state with the condition that the sudden death does not occur at or around the critical point. In Appendix B, we tackle this matter in detail.

It can be inferred that all these results  have been established based on a basic theoretical framework, simply applied to other complex models. Moreover, it can be predicted that such a framework is more likely to be upgraded to higher quenches (triple,...) to create more QCs compared to double quenches.

\textit{Acknowledgment.}$-$
H. Cheraghi thanks National Science Centre (NCN, Poland) under Grant No. 2019/35/B/ST3/03625. for the support of this work. We are also thankful to T. Mohammad Alizadeh and J. Vahedi for proofreading and editing the article.

%%%%%%%%%%%%%%%%%%%%%%%%%%%%%%%%%%%%%%%%%%%
%%%%%%%%%%%%%%%%%%%%%%%%%%%%%%%%%%%%%%%%%%%
\appendix
\section{Quantum discord}  
The total correlation can be calculated by 
\begin{eqnarray}
{\cal I}({\rho _{i,i + 1}}) = S({\rho _i}) + S({\rho _{i,i + 1}}) + \sum\limits_{\alpha  = 0}^3 {{\lambda _\alpha }} {\log _2}{\lambda _\alpha },
\end{eqnarray}
where $\lambda _\alpha$ is the eigenvalue of the density matrix $\rho _{i,i + 1}$, and
\begin{eqnarray}
S(\rho _i) =  - \sum\limits_{\xi  =  \pm 1} {\left[ {\frac{{1 + \xi {c_4}}}{2}{{\log }_2}\left( {\frac{{1 + \xi {c_4}}}{2}} \right)} \right]} .
\end{eqnarray}
Here, $c_{i=1,...,4}$, are expressed as
\begin{eqnarray}
{c_{1/2}}  &=& 2({Z_{i,i + 1}} \pm {f_{i,i + 1}}),    \nonumber \\
{c_{3/4}}  &=& X_{i,i + 1}^ +  \pm X_{i,i + 1}^ -  - Y_{i,i + 1}^ +  \mp Y_{i,i + 1}^ - .
\end{eqnarray}
Since the translational invariance of the original Hamiltonian, the single-site density matrices  $\rho _i$ and $\rho _{i,i + 1}$ are equal, we have $S(\rho _i) = S(\rho _{i,i + 1})$.

The calculation of the classical correlation ${\cal C}(\rho _{i,i + 1})$ requires an optimization over rank-1 local measurements on part $B$ of $\rho _{i,i + 1}$ (here we have taken site $j$ of $\rho _{i,j}$ as part $B$). A general set of local rank-1 measurement operators, ${B_0, B_1}$, can be defined as ${B_{k\rq{} = 0/1}} = V {\prod\nolimits_{k\rq{}}} {V^\dag } $, where $V \in U(2)$ and the projectors $\prod\nolimits_{{k\rq{}}}$are given in the computational basis $\left| 0 \right\rangle  \equiv \left|  \uparrow  \right\rangle $ and $\left| 1 \right\rangle  \equiv \left|  \downarrow  \right\rangle $. The post measurement outcomes get updated to one of the following states
\begin{eqnarray}
{\rho _{{k\rq{}}}} = \left( {\frac{1}{2} + \sum\limits_{j = 1}^3 {{\chi _{{k\rq{}}}}{S_j}} } \right) \otimes (V{\prod\nolimits_{{k\rq{}}}} {{V^\dag }} ),
\end{eqnarray}
where the elements of the density matrices are given by
\begin{eqnarray}
{\chi _{{k\rq{}}i = 1,2}}  &=& \frac{{{( - 1)^{{k\rq{}}}}{c_i}\sin \theta \cos \phi }}{{1 + {{( - 1)}^{{k\rq{}}}}{c_4}\cos \theta }},    \nonumber \\
{\chi _{{k\rq{}}3}}  &=& \frac{{{{( - 1)}^{{k\rq{}}}}{c_3}\cos \theta  + {{\mathop{\rm c}\nolimits} _4}}}{{1 + {{( - 1)}^{{k\rq{}}}}{c_4}\cos \theta }}.
\end{eqnarray}
Here, the azimuthal angle $\theta  = [0,\pi ]$ and the polar angle $\phi  = [0,2\pi ]$ represent a qubit over the Bloch sphere. By considering the normalization of the density matrices, ${\theta _{{k\rq{}}}} = {(\sum\nolimits_{j = 1}^3 {\chi _{{k\rq{}}j}^2} )^{1/2}}$, we derive the classical correlation between the spin pairs
\begin{eqnarray}
{\cal C}(\rho _{i,i + 1}) &=& Ma{x_{\{ {B_{{k\rq{}}}}\} }}\left[ {S({\rho _i}) - \frac{{S({\rho _0}) + S({\rho _1})}}{2}} \right.   \nonumber \\
 &-& \left. {c_4}\cos \theta \frac{S({\rho _0}) - S(\rho _1)}{2} \right],
\end{eqnarray}
where the von Neumann entropies are given by
\begin{eqnarray}
S(\rho _{k\rq{}}) =  - \sum\limits_{\xi  =  \pm 1} \left[ \frac{1 + \xi \theta _{k\rq{}}}{2}{{\log }_2}\left( {\frac{{1 + \xi {\theta _{k\rq{}}}}}{2}} \right) \right] .
\end{eqnarray}
It should be noted that the von Neumann entropy of $V {\prod\nolimits_{k\rq{}}} {V^\dag } $ is zero.  

%%%%%%%%%%%%%%%%%%%%%%%%%%%%%%%%%%%%%%%%%%%
%%%%%%%%%%%%%%%%%%%%%%%%%%%%%%%%%%%%%%%%%%%
\section{Nonanalyticity at the quantum critical point in  the  steady-state QCs}
A quench from a noncritical point to a critical point at the long time average of the system leads to the emergance of nonanalytical behaviors of the  steady-state QCs. It occurs when the quenched parameter as control parameter crosses the critical point, and the critical point will be far from the sudden death phenomenon. We here are going to extract the origination of these nonanalytics.

\textit{Single quench.}$-$ In general,  we assume   an initial Hamiltonian ${\cal H}(h_i)$ at $t=0$ in the equilibrium with its ground state. It undergoes a quantum quench from $h_i$ to $h_f$ that puts the system in a nonequilibrium condition. Loschmidt echo is defined as as the overlap between initial and time-evolved states, $LE(t) = |\langle {\Psi (0)|\Psi (t)} \rangle |^2$. It is demonstrated that nonequilibrium quantum phase transitions can be identified by nonanalyticities in the long-time average of the  Loschmidt echo, written as \cite{27e}
\begin{eqnarray} \label{b1}
\overline {\cal L} (h _f) = \sum\nolimits_n |\langle {E_n^f|\Psi (0)} \rangle |^4 = \sum\nolimits_n {|g_n|^4}  
\end{eqnarray}
In fact, at the critical point the sharp of $\overline {\cal L}$ changes and thus the rate function, i.e., the logarithm function of Loschmidt echo, can exactly give the quantum critical point.
With the comparison of eq.~(\ref{b1}) and eq.~(\ref{eq15}), it is easy to prove that the nonanalytical behaviors existing in $h_f=1$ are hidden in the heart of the  $g_n$ parameter where $h_f$ as the control parameter crosses the critical point.

Let us come back to the theory of the dynamical quantum phase transition \cite{27} where the Fisher zeros of Loschmidt echo are $z_l = [ \ln (\tan ^2(\Phi _k^{f})) + i\pi (2l + 1) ]/({2\varepsilon _k^{f}})$, with $l=0,1,2,...$.
As known, nonanalytic behaviors of the rate function in the form of cusps at some special times called \textit{critical times} occurs in Fisher zeros.
When $\Phi _k^{f}=\pm \pi/4$, the real parts of $z_l$ are zero, and we get the critical times $t_l = {t^*}(l + 1/2)$
with $t^* = \pi /\varepsilon _{k^*}^f$ where $k^*$ is the particular mode driven from $\cos(2\Phi _{k^*}^f) = 0$. 
 If the pre-quenched Hamiltonian is in noncritical point but the post-quenched Hamiltonian is in critical, there is no dynamical quantum phase transition in finite time \cite{58}.  In this case, $k^*=\pm \pi$ is also the gap-closing point of the post-quenched Hamiltonian, i.e., $\varepsilon _{k^*=\pm \pi}^f=0$. Instead, in this situation $t^* = \infty$. At the long time average of the system, the nonanalytic point as a cusp appears. Similarity,  we should  accept the nonanalytics of the  steady-state QCs of the system after a long-time evolution. In addition, $k^*=\pm \pi$  leads to the singularity of the Bogoliubov angle at the gap-closing momentum \cite{58}. In a sense, all  $g_n$  can play a role  to create the nonanalyticity at the critical point.

\textit{Double quench.}$-$ A double quench is defined as a quench from $h_i$ to $h_m$ and then at $T$, from $h_m$ to $h_f$. For the single quench $h_i \to h_m$  at $t=T$   we have
\begin{eqnarray}
| {\Psi (T)} \rangle  = e^{ - iTH^m}\left| \Psi (0) \right\rangle  = \sum\nolimits_{p = 0} {g_p} e^{ - iTE_p^m}| {E_p^m} \rangle   \nonumber\\
\end{eqnarray}
where $g_p = \left\langle {E_p^m|\Psi (0)} \right\rangle $. Now,  the second quench for $t>T$  is given by
\begin{eqnarray}
\left| {\Psi (t)} \right\rangle  &=& {e^{ - i(t - T){H^f}}}\left| {\Psi (T)} \right\rangle    \nonumber\\
 &=& \sum\nolimits_{p,q = 0} {{g_p}} {d_q}{e^{ - iT(E_p^m - E_q^f)}}{e^{ - itE_q^f}}| {E_q^f} \rangle   \nonumber\\
\end{eqnarray}
where $d_q = \langle {E_q^f|E_p^m} \rangle $. 
So, the long-time average of the  Loschmidt echo and the concurrence are as the following
\begin{eqnarray} 
\overline {\cal L} (h _f) &=&  \sum\nolimits_{p,q,w = 0} \left| g_p \right| ^2\left| g_w \right|^2\left| d_q \right|^2\left| d_w \right|^2     \nonumber\\
\langle C \rangle  &=& \sum\nolimits_{p,q = 0} \left| g_p \right| ^2\left| d_q \right|^2  C_{q,q}
\end{eqnarray}
where $C_{q,q} = \langle E_q^f|C|E_q^f \rangle $. By comparing these equations one concludes that the nonanalyticity in Loschmidt echo directly is related to $ g_p,  g_w, d_q$ and $d_w$ which are parameters that construct the  concurrence. For this case, the critical times arisen from zeros of Loschmidt echo are  $t_l^* = \frac{\pi }{{2\varepsilon _{{k^*}}^f}}\left[ {(2l + 1) - {\varphi _{k^*}}} \right]$, with $l=0,1,2,...$, where $\varphi _{k^*}$ is a phase shift originated from the time evolution for $t < T$  \cite{59}. Here, for $h_f=1.0$ we infer that  $k^*=\pm \pi$  is  the gap-closing point of the post-quenched Hamiltonian, and hence $\varepsilon _{k^*=\pm \pi}^f=0$. Therefore, we extract the same results as what we have for the single quench, unveiling the nonanalyticity at the steady-state QCs.

%%%%%%%%%%%%%%%%%%%%%%%%%%%%%
%%%%%%%%%%%%%%%%%%%%%%%%%%%%%
%%%%%%%%%%%%%%%%%%%%%%%%%%%%%
\bibliographystyle{plain}

\end{document}